\documentclass[12pt]{article}
 \usepackage{amssymb}
 \usepackage{latexsym}
\usepackage{graphicx}
\usepackage{epstopdf}

\begin{document}
\begin{titlepage}

\vskip 5em

\begin{center}
{\bf \huge
Flavoured Large $N$ Gauge Theory in an External Magnetic Field}
\vskip 3em

\centerline{\bf Veselin G. Filev${}^{\star}$, Clifford V. Johnson${}^{\star}$, R. C. Rashkov${}^{\dagger}$\footnote{On leave from Dept of Physics, Sofia University, Bulgaria.} and K. S. Viswanathan${}^{\dagger}$}

\bigskip

  \centerline{${}^{\star}$\it Department of Physics and Astronomy, University of
Southern California}
\centerline{\it Los Angeles, CA 90089-0484, U.S.A.}

\centerline{\small \tt  filev@usc.edu, johnson1@usc.edu}
\bigskip
\centerline{${}^{\dagger}$\it Department of Physics Simon Fraser University and IRMACS Centre}
\centerline{\it Burnaby, BC, V5A 1S6, Canada}

\centerline{\small \tt  kviswana@sfu.ca; rash@phys.uni-sofia.bg}

\bigskip
\bigskip

\vskip 4em
\begin{abstract}
We consider a D7-brane probe of AdS$_{5}\times S^5$ in the presence of pure gauge $B$-field. In the dual gauge theory, the $B$-field couples to the fundamental matter introduced by the D7-brane and acts as an external magnetic field. The $B$-field supports a 6-form Ramond-Ramond potential on the D7-branes world volume that breaks the supersymmetry and enables the dual gauge theory to develop a non-zero fermionic condensate. We explore the dependence of the fermionic condensate on the bare quark mass $m_{q}$ and show that at zero bare quark mass a chiral symmetry is spontaneously broken. A study of the meson spectrum reveals a coupling between the vector and scalar modes, and in the limit of weak magnetic field we observe Zeeman splitting of the states. We also observe the characteristic $\sqrt{m_{q}}$ dependence of the ground state corresponding to the Goldstone boson of spontaneously broken chiral symmetry.
\end{abstract}
\end{center}
\end{titlepage}

\section{Introduction}

    In  recent years, progress has been made in the study of  gauge theory with matter in the fundamental representation in the context of gauge/string dualities generalizing the AdS/CFT correspondence. One way
    to achieve this is by introducing D7-branes in the probe limit\cite{Karch:2002sh} that amounts to  the condition $N_{f} \ll N_{c}$. The fundamental strings stretched between the stack of $N_{c}$ D3-branes and
     the $N_{f}$ flavor D7-branes give rise to $\cal N$=2 hypermultiplet. The separation of the D3- and D7-branes in the transverse
     directions corresponds to the mass of the hypermultiplet, the classical shape of the D7-brane encodes the value of the fermionic
     condensate, and its fluctuations describe the light meson spectrum of the theory\cite{Kruczenski:2003be}. This technique for
      introducing fundamental matter has been widely employed in different backgrounds. Of particular interest is the study of non
      supersymmetric backgrounds and phenomena such as spontaneous chiral symmetry breaking. These phenomena were first studied in this context in ref. \cite{Babington:2003vm}, using analytical and  numerical techniques. In  several works this approach was further developed, and has proven itself a powerful tool for the exploration of gauge theories, in particular, for the description of their thermodynamic properties or for the building of phenomenological models relevant to QCD \cite{Kruczenski:2003uq}--\cite{Hoyos:2006gb}.

In this paper we will be interested in introducing fundamental matter  into the gauge theory  in the presence of an external
   electromagnetic field that couples to the fundamental fermions. The  supersymmetry will be explicitly broken by the external field, and we will observe spontaneous symmetry breaking, and  non--trivial mixing in the spectrum of mesons.

\section{Fundamental Matter in an External Magnetic Field }
\subsection{Basic Configuration}

Let us consider the AdS$_{5} \times S^5$ geometry describing the near-horizon physics of a collection of $N_{c}$ extremal D3-branes.
\begin{eqnarray}
ds^2&=&\frac{u^2}{R^2}(-dx_{0}^2+dx_1^2+dx_2^2+dx_3^2)+R^2\frac{du^2}{u^2}+R^2d\Omega_{5}^2\ ,\label{AdS}\\
g_{s}C_{(4)}&=&\frac{u^4}{R^4}dx^0\wedge dx^1\wedge dx^2 \wedge dx^3\ ,\nonumber\\\
e^\Phi&=&g_s\ ,\nonumber\\
R^4&=&4\pi g_{s}N_{c}\alpha'^2\ ,\nonumber
\end{eqnarray}
Where $d\Omega_5^2$ is the unit metric on a round $S^5$. In order to introduce fundamental matter we first rewrite the metric in the following form, with $d\Omega_3^2$ the metric on a unit $S^3$:
\begin{eqnarray}
ds^2&=&\frac{\rho^2+L^2}{R^2}[ - dx_0^2 + dx_1^2 +dx_2^2 + dx_3^2 ]+\frac{R^2}{\rho^2+L^2}[d\rho^2+\rho^2d\Omega_{3}^2+dL^2+L^2d\phi^2],\nonumber\\
d\Omega_{3}^2&=&d\psi^2+\cos^2\psi d\beta^2+\sin^2\psi d\gamma^2, \label{geometry1}
\end{eqnarray}
where $\rho, \psi, \beta,\gamma$ and $L,\phi$ are polar coordinates in the transverse $\mathbb{R}^4$ and $\mathbb{R}^2$ respectively. Note that: $u^2=\rho^2+L^2$.
We use $x_{0},x_{1},x_{2},x_{3},\rho,\psi,\beta,\gamma$ to parametrise the world volume of the D7-brane and consider the following ansatz \cite{Kruczenski:2003be} for its embedding:
\begin{eqnarray}
\phi\equiv {\rm const},\quad L\equiv L(\rho)\nonumber \label{anzatsEmb},
\end{eqnarray}
leading to the following form of the induced metric on its worldvolume:
\begin{equation}
d\tilde s=\frac{\rho^2+L(\rho)^2}{R^2}[ - dx_0^2 + dx_1^2 +dx_2^2
+dx_3^2]+\frac{R^2}{\rho^2+L(\rho)^2}[(1+L'(\rho)^2)d\rho^2+\rho^2d\Omega_{3}^2] \  .
\label{inducedMetric}
\end{equation}
Now let us consider the general DBI action:
\begin{eqnarray}
S_{DBI}=-\mu_{7}\int\limits_{{\cal M}_{8}}d^{8}\xi e^{-\Phi}[-{\rm det}(G_{ab}+B_{ab}+2\pi\alpha' F_{ab})]^{1/2}\  . \label{DBI}
\end{eqnarray}

Here $\mu_{7}=[(2\pi)^7\alpha'^4]^{-1}$ is the D7-brane tension, $G_{ab}$ and $B_{ab}$ are the induced metric and $B$-field on the D7-brane's world volume, while $F_{ab}$ is its world--volume gauge field. A simple way to
introduce  magnetic field would be to consider a pure gauge $B$-field along parts of the D3-branes' world volume, {\it e.g.}:
\begin{equation}
B^{(2)}= Hdx_{2}\wedge dx_{3} \label{anzats}\ .
\end{equation}
Since $B_{ab}$ can be mixed with the gauge field strength $F_{ab}$, this is equivalent to a magnetic field on the world--volume.
Recently a similar approach was used to study drag force in SYM plasma \cite{Matsuo:2006ws}. Note that since the $B$-field is pure gauge, $dB=0$, the corresponding background is still a solution to the supergravity equations of motion.
On the other hand, the gauge field $F_{ab}$ comes at next order in the $\alpha'$ expansion compared to the metric and the $B$-field components. Therefore
to study the classical embedding of the D-brane one can study only the $(G_{ab}+B_{ab})$ part of the DBI-action. However, because of the
presence of  the $B$-field, there will be terms ot first order in $\alpha'$ in the full action linear in the gauge field $F_{ab}$. Hence  integrating out $F_{ab}$ will result in a constraint for the classical embedding of the D7-brane.

Since for our configuration, we have  that:
\begin{eqnarray*}
B^{(2)}\wedge B^{(2)}=0\ ,\quad B^{(2)}\wedge C_{(4)}=0\ ,
\end{eqnarray*}
and at first order in $\alpha'$
the only contribution to the Wess-Zummino is
\begin{eqnarray}
2\pi\alpha'\mu_{7}\int F_{(2)}\wedge C_{(6)}\ .
\label{potentials}
\end{eqnarray}

By using the following  expansion in the DBI action:
\begin{equation}
[-{\rm det}(E_{ab}+2\pi\alpha'F_{ab})]^{1/2}=\sqrt{E}+\pi\alpha'\sqrt{E}E^{ba}F_{ab}+O(F^2);~~~E=-{\rm det}E_{ab};\ ,
\end{equation}
where we have introduced $E_{ab}=G_{ab}+B_{ab}$ as a notation for the generalized induced metric,  we obtain the following action to first order in $\alpha'$:
\begin{equation}
S_{F}=\pi\alpha'\frac{\mu_7}{g_s}\int\limits_{{\cal M}_{8}}d^8\xi\sqrt{E}E^{[ab]}F_{[ab]}+2\pi\alpha'\mu_7\int F_{(2)}\wedge C_{(6)}\ .
\end{equation}
The resulting equation of motion does not contain $A_a$ and sets the following constraint for the $C_{(6)}$ potential induced by the gauge $B$-field.
\begin{equation}
\frac{g_s}{6!}\epsilon^{ab\tilde\mu_{1}\dots\tilde\mu_{6}}\partial_{a}C_{\tilde\mu_{1}\dots\tilde\mu_{6}}=-\partial_{a}(\sqrt{E}E^{[ba]});~~~a,b,\tilde\mu_{1},\dots\tilde\mu_{6}\in{\cal M}_{8};\ .
\label{cosntr}
\end{equation}
Note that  $C_{(6)}$ has a dynamical term proportional to $1/\kappa_{0}^2$  in the supergravity action, and that the D7-brane action is proportional to $\mu_{7}=2\pi/\kappa_{0}^2$. Therefore they are
at the same order in~$\alpha'$ \cite{Johnson:2003gi}. We must solve for $C_{(6)}$ using the action:
\begin{equation}
S_{C_{(6)}}=\mu_{7}\int B_{(2)}\wedge C_{(6)}-\frac{1}{4k_{0}^2}\int d^{10}x\sqrt{-G}|dC_{(6)}|^2\label{dynamics}\ .
\end{equation}
The solution obtained from equation (\ref{dynamics}) has to satisfy the constraint given in equation~(\ref{cosntr}). Our next goal will be to find a consistent
ansatz for $C_{(6)}$. To do this let us consider the classical contribution to the DBI action:
\begin{eqnarray}
S_{NS}=-\frac{\mu_{7}}{g_s}\int d^8\xi\sqrt{E}\label{decopl}\ .
\end{eqnarray}
From equation (\ref{decopl}) one can solve for the classical embedding of the D7-brane, which amounts to second order differential
equation for $L(\rho)$ with some appropriate solution $L_0(\rho)$. After substituting $L_0(\rho)$ in (\ref{decopl}) we can extract the form of the $C_{(6)}$ potential induced by the $B$-field. However one still has to satisfy the constraint (\ref{cosntr}).
It can be verified that with the choice (\ref{anzats}) for the $B$-field and the ansatz of equation  (\ref{inducedMetric}) for the induced metric, the right-hand side of equation (\ref{cosntr}) is zero. Then equation (\ref{cosntr}) and the effective action (\ref{dynamics}) boil down to finding a consistent ansatz for $C_{(6)}$ satisfying:
\begin{eqnarray}
&\partial_{\mu}(\sqrt{-G}dC_{6}^{\mu01\rho\psi\alpha\beta})&=-\frac{\mu_{7}\kappa_{0}^2}{\pi}H\delta(L-L_0 (\rho))\ ,\\
{\rm or}\qquad &\partial_{\mu}(\sqrt{-G}dC_{6}^{\mu01L\psi\alpha\beta})&=- L_0' (\rho)\frac{\mu_{7}\kappa_{0}^2}{\pi}H\delta(L-L_0 (\rho))\ ,\\
&\epsilon^{ab\tilde\mu_{1}\dots\tilde\mu_{6}}\partial_{a}C_{\tilde\mu_{1}\dots\tilde\mu_{6}}&=0;~~~a,b,\tilde\mu_{1},\dots\tilde\mu_{6}\in{\cal M}_{8};\ .
\end{eqnarray}
%
One can verify that the choice:
\begin{eqnarray}
{C_{(6)}}_{01\rho\psi\alpha\beta}=\frac{1}{7}f(\rho,L,\psi),\quad {dC_{(6)}}_{L01\rho\psi\alpha\beta}=\partial_{L}f\ ,
\end{eqnarray}
is a consistent ansatz and the solution for the $C_{(6)}$ field strength can be found to be:
\begin{equation}
{dC_{(6)}}_{L01\rho\psi\alpha\beta}=\frac{\mu_{7}\kappa_{0}^2}{\pi}H\frac{\rho^3R^4}{L(\rho^2+L^2)^2}\Theta(L-L_0(\rho))\sin\psi\cos\psi\ .
\label{closed form}
\end{equation}

It is this potential which breaks the supersymmety. It is important to note that there is no contradiction between the fact that the $B$--field that we have chosen does not break the supersymmetry of the AdS$_5\times S^5$ supergravity background, on the one hand, and the fact that the physics of the D7--brane probing that background does have supersymmetry broken by the $B$--field, on the other. This is because the physics of the probe does not back--react on the geometry.

In what follows, we will study the physics  of the D7-branes and the resulting  dual gauge theory physics. Among the solutions for the D7-brane embedding, there will be a class with non-trivial profile having zero asymptotic separation between the D3- and D7-branes. This corresponds to a non-zero fermionic condensate at zero bare quark mass. Therefore the non-zero background magnetic field will spontaneously break the chiral symmetry.  Geometrically this corresponds to breaking of the $SO(2)$ rotational symmetry in the $(L, \phi)$-plane \cite{Kruczenski:2003be}.

\subsection{Properties of the Solution}
We now proceed with the exploration of the properties of the classical D7-brane embedding.
If we consider the action (\ref{decopl}) at leading order in $\alpha'$, we get the following effective lagrangian:
\begin{equation}
{\cal L}=-\frac{\mu_{7}}{g_s}\rho^3\sin\psi\cos\psi\sqrt{1+L'^2}\sqrt{1+\frac{R^4H^2}{(\rho^2+L^2)^2}}\ .
\label{lagrangian}
\end{equation}
The equation of motion for the profile $L_0(\rho)$ of the D7-brane is given by:
\begin{equation}
\partial_{\rho}\left(\rho^3\frac{L_0'}{\sqrt{1+L_0'^2}}\sqrt{1+\frac{R^4H^2}{(\rho^2+L_0^2)^2}}\right)+ \frac{\sqrt{1+L_0'^2}}{\sqrt{1+\frac{R^4h^2}{(\rho^2+L_0^2)^2}}}\frac{2\rho^3L_0R^4H^2}{(\rho^2+L_0^2)^3}=0\ .
\label{eqnMnL}
\end{equation}
As expected for large $(L_0^2+\rho^2) \to \infty$ or $H \to 0$, we get the equation for the pure AdS$_{5}\times S^5$ background \cite{Karch:2002sh}:
\begin{eqnarray*}
\partial_{\rho}\left(\rho^3\frac{L_0'}{\sqrt{1+L_0'^2}}\right)=0\ .
\end{eqnarray*}
Therefore the solutions to equation (\ref{eqnMnL}) have the following behavior at infinity:
\begin{equation}
L_0(\rho)=m+\frac{c}{\rho^2}+\dots.
\end{equation}
where the parameters $m$ (the asymptotic separation of the D7- and D3- branes) and $c$ (the degree of bending of the D7-brane) are related to the bare quark mass $m_{q}=m/2\pi\alpha'$ and the fermionic condensate $\langle\bar\psi\psi\rangle\propto -c$ respectively \cite{Kruczenski:2003uq}. As we shall see below, the presence of the external magnetic field and its effect on the dual SYM  provide a non vanishing value for the   fermionic condensate, furthermore the theory exhibits chiral symmetry breaking.

   Now notice that $H$ enters in (\ref{lagrangian}) only through the combination $H^2R^4$. The other natural scale is the asymptotic separation $m$. It turns out that different physical configurations can be studied in terms of  the ratio $\tilde m^2={m^2}/{(H R^2)}$: Once the $\tilde m$ dependence of our solutions  are known, the $m$ and $H$ dependence follows. Indeed let us introduce dimensionless variables {\it via}:
  \begin{eqnarray}
  \rho=R\sqrt{H}\tilde\rho\ , \quad
  L_0=R\sqrt{H}\tilde L\ , \quad
  L_0'(\rho)=\tilde L'(\tilde\rho)\ .\label{cordchange}
  \end{eqnarray}
The equation of motion (\ref{eqnMnL})  then takes the form:
\begin{equation}
\partial_{\tilde\rho}\left(\tilde\rho^3\frac{\tilde L'}{\sqrt{1+{\tilde L}'^2}}\sqrt{1+\frac{1}{(\tilde\rho^2+\tilde L^2)^2}}\right)+ \frac{\sqrt{1+\tilde L'^2}}{\sqrt{1+\frac{1}{(\tilde\rho^2+\tilde L^2)^2}}}\frac{2\tilde\rho^3\tilde L}{(\tilde\rho^2+\tilde L^2)^3}=0
\label{eqnMnLD}
\end{equation}
The solutions for $\tilde L(\tilde\rho)$ can be expanded again to:
\begin{equation}
\tilde L(\tilde\rho)=\tilde m+\frac{\tilde c}{\tilde\rho^2}+\dots, \label{ExpansionD}
\end{equation}
and using the transformation (\ref{cordchange}) we can get:
\begin{equation}
c=\tilde c R^3H^{3/2} \label{Hdepend}\ .
\end{equation}

It is instructive to study first the properties of (\ref{eqnMnLD}) for $\tilde m\gg1$, which corresponds to weak magnetic field $H\ll{m^2}/{R^2}$,
  or equivalently large quark mass $m\gg R\sqrt{H}$.

\subsubsection{Weak Magnetic Field}

In order to analyze the case of weak magnetic field let us expand $\tilde L(\tilde\rho)=\tilde m+\eta(\tilde\rho)$ and linearize equation (\ref{eqnMnLD})
 while leaving only the leading terms in $(\tilde\rho^2+\tilde m^2)^{-1}$. The result is:
\begin{equation}
\partial_{\tilde\rho}\left(\tilde\rho^3\eta'\right)+\frac{2\tilde\rho^3\tilde m}{(\tilde\rho^2+\tilde m^2)^3}=0\ ,
\label{eqnSmA}
\end{equation}
which has the general solution:
\begin{equation}
\eta(\tilde\rho)=\frac{C_{1}}{\tilde\rho^2}-\frac{\tilde m}{4\tilde\rho^2(\tilde m^2+\tilde\rho^2)}+C_{2}\ . \label{solutionSmA}
\end{equation}
From the definition of $\eta(\tilde\rho)$ and equation~(\ref{ExpansionD}) we can see that $C_{1}=\tilde c$ and since $\eta|_{\tilde\rho\to\infty}=0$ we have
 $C_{2}=0$. Now if we consider $\tilde m$ large enough, equation (\ref{solutionSmA}) should be valid for all $\tilde\rho$. It turns out that if we require
 that our solution be finite as $\tilde\rho \to 0$ we can determine the large~$\tilde m$ behavior of $\tilde c$. Indeed the second term in (\ref{solutionSmA})
 has the expansion:
\begin{equation}
-\frac{\tilde m}{4\tilde\rho^2(\tilde m^2+\tilde\rho^2)}=-\frac{1}{4\tilde m}\frac{1}{\tilde\rho^2}+\frac{1}{4\tilde m^3}+O(\tilde\rho^2)\ .
\end{equation}
Therefore we deduce that:
\begin{equation}
C_{1}=\tilde c=\frac{1}{4\tilde m} \label{1/m}\ ,
\end{equation}
and finally, we get for the profile of the D7-brane for $\tilde m\gg 1$:
\begin{equation}
\tilde L(\tilde\rho)=\tilde m+\frac{1}{4\tilde m}\frac{1}{\tilde\rho^2}-\frac{\tilde m}{4\tilde\rho^2(\tilde m^2+\tilde\rho^2)}\ . \label{embSmA}
\end{equation}
If we go back to dimensionful parameters we can see, using equations (\ref{Hdepend}) and (\ref{1/m}) that for weak magnetic field $H$ the theory has developed a
 fermionic condensate:
\begin{equation}
\langle\bar\psi\psi\rangle \propto -c =-\frac{R^4}{4m}H^2\ . \label{condSmA}
\end{equation}

However this formula is valid only for sufficiently large $m$ and we cannot make any prediction for the value of the fermionic condensate at zero
quark mass. To go further, the involved form of equation (\ref{eqnMnLD}) suggests the use of numerical techniques.

\subsubsection{Numerical Results}

\begin{figure}[h] 
   \centering
   \includegraphics[width=10cm]{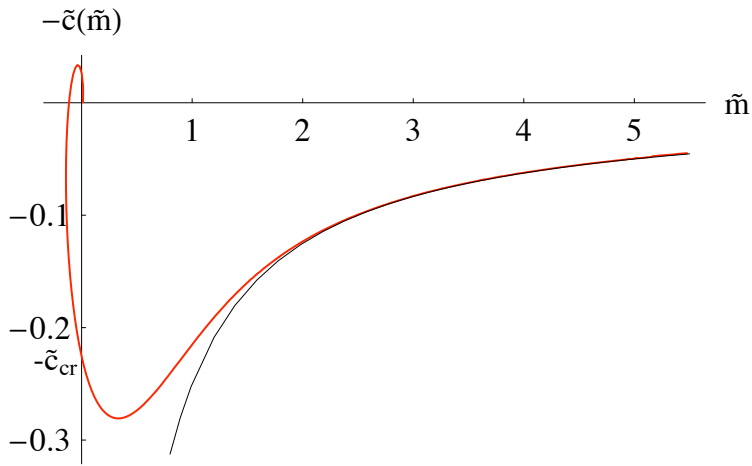}
   \caption{\small The black line corresponds to (\ref{1/m}), one can observe that the analytic result is valid for large $\tilde m$.
    It is also evident that for $\tilde m=0$ $\langle\bar\psi\psi\rangle\neq0$. The corresponding value of the condensate is $\tilde c_{\rm cr}=0.226$}.   \label{fig:fig1}
\end{figure}

In this subsection we solve numerically equation (\ref{eqnMnLD}) for the embedding of the D7-brane, using Mathematica. It is convenient
 to use initial conditions in the IR as has been recently discussed in the literature \cite{Albash:2006ew}, \cite{Albash:2006bs}.We use the boundary condition  $\tilde L'(\tilde\rho)\vert_{\tilde\rho=0}=0$. We used  shooting
 techniques to generate the embedding of the D7 for a wide range of $\tilde m$. Having done so we expanded numerically
 the solutions for $\tilde L(\tilde\rho)$ as in equation (\ref{ExpansionD}) and generated the points in the  $(\tilde m,-\tilde c)$ plane corresponding to the solutions. The resulting plot is presented in figure~\ref{fig:fig1}.

As one can see there is a non zero fermionic condensate for zero bare quark mass, the corresponding value of the condensate is $\tilde c_{\rm cr}=0.226$. It is also evident that the analytical expression for the condensate (\ref{1/m}) that we got in the previous section is valid for large $\tilde m$, as expected. Now using equation (\ref{Hdepend}) we can deduce the dependence of $c_{\rm cr}$ on $H$:
\begin{equation}
c_{\rm cr}=\tilde c_{\rm cr}R^3H^{3/2}=0.226R^3H^{3/2}\ . \label{Ccr}
\end{equation}

\begin{figure}[h] 
   \centering
   \includegraphics[width=9cm]{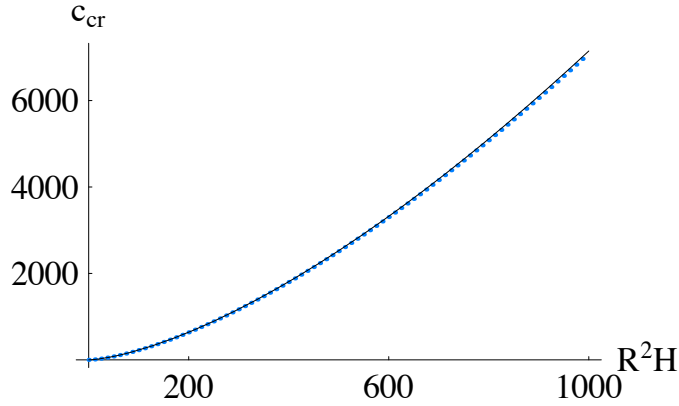}
   \caption{\small A plot of the magnitude of the fermionic condensate at zero bare quark mass $c_{\rm cr}$ as function of $R^2H$, the black curve represents
   equation (\ref{Ccr}).}
   \label{fig:fig2}
\end{figure}

It is interesting to check the consistency of our numerical analysis by solving equation~(\ref{eqnMnL}) numerically and extracting the value of $c_{\rm cr}$
 for wide range of $R^2H$, the resulting plot fitted with equation (\ref{Ccr}) is presented in figure \ref{fig:fig2}.

\begin{figure}[h] 
   \centering
   \includegraphics[width=9cm]{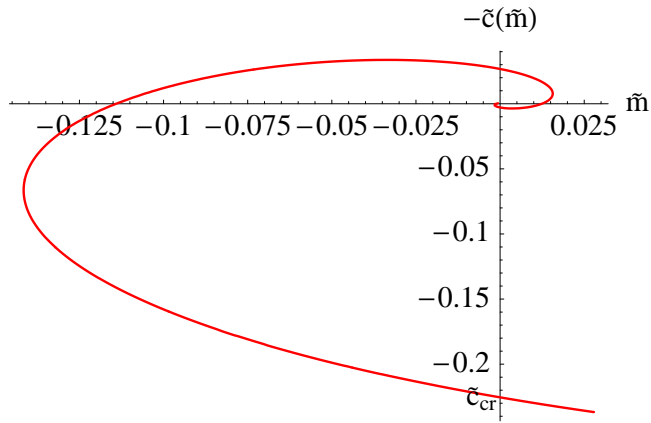}
   \caption{\small A magnification of figure \ref{fig:fig1} to show the spiral behavior near the origin of the $(-\tilde c,\tilde m)$-plane.}
   \label{fig:spiral}
\end{figure}

  Another interesting feature of our
   phase diagram is the spiral behavior near the origin of the $(\tilde m,-\tilde c)$-plane which can be seen in figure \ref{fig:spiral}. A similar feature has been observed
  in ref.~\cite{Albash:2006bs}, where the authors have argued that only the lowest branch of the spiral corresponding to positive values of
   $m$ is the stable one (corresponding to the lowest energy state).  The spiral behavior near the origin signals instability of the
   embedding corresponding to $L_0\equiv 0$. If we trace the curve of the diagram in
   figure \ref {fig:spiral} starting from large $m$, as we go to smaller values of $m$ we will reach zero bare quark mass for some
   large negative value of the fermionic condensate $c_{cr}$. Now if we continue tracing along  the diagram one can verify numerically that all other points correspond to embeddings of the D7-brane which intersect the D3-brane at least once. (Note also that in ref.~\cite{Babington:2003vm}, such behavior was considered inconsistent  with the interpretation of the embedding as a re-normalization group flow.) After further study one finds  that the part of the diagram corresponding to negative values of $\tilde m$ represents solutions for the D7-brane embedding which intersect the D3-branes odd number of times, while the positive part of the spiral represents solutions which intersect the D3-branes even number of times. The lowest positive branch corresponds to solutions which don't intersect the D3-branes and is the stable one, while the upper branches have correspondingly $2,4, {\it etc.,}$ intersection points and are ruled out.

\section{Meson Spectrum}
\subsection{General Properties}
We study the scalar meson spectrum. To do so we will consider quadratic fluctuations \cite{Kruczenski:2003be} of the embedding of the D7-brane in the
transverse $(L,\phi)$-plane. It can be shown that because of the diagonal form of the metric the fluctuation modes along the $\phi$ coordinate decouple
from the one along $L$. However, because of the non-commutativity introduced by the $B$-field we may expect the scalar fluctuations to couple to the vector
fluctuations. This has been observed in ref. \cite{Arean:2005ar}, where the authors considered the geometric dual to non-commutative super Yang Mills . In our case the mixing will be even stronger,
because of the non-trivial profile for the D7-brane embedding, resulting from the broken supersymmetry.

Let's proceed with obtaining the action for the fluctuations. To obtain the contribution from the DBI part of the action we consider the expansion:
\begin{eqnarray}
L=L_0(\rho)+2\pi\alpha'\chi,\label{fluct}\quad\phi=0+2\pi\alpha'\ ,
\end{eqnarray}
where $L_0(\rho)$ is the classical embedding of the D7-brane solution to equation (\ref{eqnMnL}). To second order in $\alpha'$ we have the following expression:
\begin{equation}
E_{ab}=E^{0}_{ab}+2\pi\alpha'E^{1}_{ab}+(2\pi\alpha')^2E^{2}_{ab}\ ,
\label{33}
\end{equation}
where $E^0,E^1,E^2$ are given by:
\begin{eqnarray}
&E^{0}_{ab}&=G_{ab}(\rho,L_0(\rho),\psi)+B_{ab},\nonumber\\
&E^{1}_{ab}&=\frac{R^2{L_0}'}{\rho^2+L_0^2}\left(\partial_a\chi\delta_{b}^{\rho}+\partial_b\chi\delta_{a}^{\rho}\right)+\partial_{L_0}G_{ab}\chi+F_{ab}\label{ExpnMetrc}\\
&E^{2}_{ab}&=\frac{R^2}{\rho^2+L_0^2}\left(\partial_{a}\chi\partial_{b}\chi+L_0^2\partial_{a}\Phi\partial_{b}\Phi\right)-\frac{2R^2L_0L_0'}{(\rho^2+L_0^2)^2}\left(\partial_{a}\chi\delta_{b}^{\rho}+\partial_{b}\chi\delta_{a}^{\rho}\right)\chi+\frac{1}{2}\partial_{L_0}^{2}G_{ab}\chi^2\ .\nonumber
\end{eqnarray}
Here $G_{ab}$ and $B_{ab}$ are the induced metric and B field on the D7-brane's world volume. Now we can substitute equation (\ref{ExpnMetrc}) into equation (\ref{decopl}) and expand to second order in $\alpha'$. It is convenient \cite{Arean:2005ar} to introduce
the following matrices:

\begin{equation}
||{E_{ab}^0}||^{-1}=S+J,
\end{equation}
where $S$ is diagonal and $J$ is antisymmetric:
\begin{eqnarray}
&||S^{ab}||&={\rm diag}\{-G_{11}^{-1},G_{11}^{-1},\frac{G_{11}}{G_{11}^{2}+H^2},\frac{G_{11}}{G_{11}^{2}+H^2},G_{\rho\rho}^{-1},G_{\psi\psi}^{-1},G_{\alpha\alpha}^{-1},G_{\beta\beta}^{-1}\}\ ,\label{S}\\
&J^{ab}&=\frac{H}{G_{11}^{2}+H^2}(\delta_{3}^{a}\delta_{2}^{b}-\delta_{3}^{b}\delta_{2}^{a})\ ,\label{J}\\
&G_{11}&=\frac{\rho^2+{L_0}^2}{R^2}\ ;~~~G_{\rho\rho}=R^2\frac{(1+{L'_0}^2)}{\rho^2+L_0^2}\ ;~~~G_{\psi\psi}=\frac{R^2\rho^2}{\rho^2+L_0^2};\nonumber\\
&G_{\alpha\alpha}&=\cos^2\psi G_{\psi\psi}\ ;~~~G_{\beta\beta}=\sin^2\psi G_{\psi\psi}\ .
\end{eqnarray}

Now it is straightforward to get the effective action. At first order in $\alpha'$ the action for the scalar fluctuations is the first variation of
the classical action (\ref{decopl}) and is satisfied by the classical equations of motion. The equation of motion for the gauge field at first order
was considered in Section 2 for the computation of the $C_{(6)}$ potential induced by the $B$- field. Therefore we focus on the second order contribution from the DBI action.

After integrating by parts and taking advantage of the Bianchi identities for the gauge field, we end up with the following terms.
For $\chi$:
\begin{eqnarray}
{\cal L_{\chi}} \propto \frac{1}{2}\sqrt{-E^0}\frac{R^2}{\rho^2+{L_0}^2}\frac{S^{ab}}{1+{L'_0}^2}\partial_{a}\chi\partial_{b}\chi+\left[\partial_{L_0}^2\sqrt{-E^0}-\partial_\rho\left(\partial_{L_0}\sqrt{-E^0}\frac{L'_0}{1+{L'_0}^2}\right)\right]\frac{1}{2}\chi^2\ ,
\label{Schi}
\end{eqnarray}
and for $F$:
\begin{eqnarray}
{\cal L}_{F} \propto\frac{1}{4}\sqrt{-E^0}S^{aa'}S^{bb'}F_{ab}F_{a'b'}\ ,
\label{SF}
\end{eqnarray}
and the mixed $\chi$--$F$ terms:
\begin{eqnarray}
&{\cal L}_{F\chi}&\propto \frac{\sin2\psi}{2}f\chi F_{23}\ ,
\label{SFchi}
\end{eqnarray}
and for $\Phi$:
\begin{equation}
{\cal L}_{\Phi}\propto\frac{1}{2}\sqrt{-E^0}\frac{R^2{L_0}^2}{\rho^2+L_0^2} S^{ab}\partial_{a}\Phi\partial_{b}\Phi\ ,
\label{Sphi}
\end{equation}
where the function $f$ in (\ref{SFchi}) is given by:
\begin{eqnarray}
&f(\rho)&=\partial_{\rho}\left(g(\rho)\frac{L'_0}{1+{L_0}'^{2}}J^{23}\right)+J^{32}\partial_{L_0}g(\rho)+2g(\rho)J^{23}S^{22}\partial_{L_0}G_{11}\ ,\\
{\rm with}\quad&g(\rho)&=\frac{\sqrt{-E^0}}{\sin\psi\cos\psi}=\rho^3\sqrt{1+{L_0}'^2}\sqrt{1+\frac{R^4H^2}{(\rho^2+L_0^2)^2}}\ .\nonumber
\label{Fchi}
\end{eqnarray}

As can be seen from equation (\ref{SFchi}) the $A_2,A_3$ components of the gauge field couple to the scalar field $\chi$ via the function $f$. Note that since
for $\rho \to \infty$ and $L\to\infty$, we see that $J^{23}\to 0$, the mixing of the scalar and vector field decouples asymptoticly. In order
to proceed with the analysis we need to take into account the contribution from the Wess-Zumino part of the action. The relevant terms to second order
in $\alpha'$ are \cite{Arean:2005ar}:
\begin{equation}
S_{WZ}=\frac{(2\pi\alpha')^2}{2}\mu_{7}\int{F_{(2)}\wedge F_{(2)}\wedge C_{(4)}}+(2\pi\alpha')\mu_{7}\int F_{(2)}\wedge B_{(2)}\wedge \tilde P[C_{(4)}]\ ,
\label{WZ}
\end{equation}
where $C_{(4)}$ is the background R-R potential given in equation~(\ref{AdS}) and $\tilde C_{(4)}$ is the pull back of its magnetic dual. One can show
that:
\begin{equation}
\tilde C_{4}=\frac{R^4}{g_{s}}\frac{2\rho^2+L^2}{(\rho^2+L^2)^2}L^2 \sin\psi\cos\psi d\psi\wedge d\alpha\wedge d\beta\wedge d\phi\ .
\end{equation}
Writing $\phi=2\pi\alpha'\Phi$  we write for the pull back $P[\tilde C_{(4)}]$:
\begin{equation}
P[\tilde C_{(4)}]=-\frac{2\pi\alpha'}{g_s}\frac{\sin2\psi}{2}K(\rho)\partial_{a}\Phi d\psi\wedge d\alpha\wedge d\beta\wedge dx^a,
\label{pulC}
\end{equation}
where we have defined:
\begin{equation}
K(\rho)=-R^4L_0^2\frac{2\rho^2+{L_0}^2}{(\rho^2+L_0^2)^2}
\end{equation}
Now note that the $B$-field has components only along $x^2$ and $x^3$, therefore $dx^a$ in equation (\ref{pulC}) can be only $d\rho,dx^0$ or $dx^1$.
This will determine the components of the gauge field which can mix with $\Phi$, However after integrating by parts and using the Bianchi identities
one can get the following simple expression for the mixing term:
\begin{equation}
-(2\pi\alpha')^2\frac{\mu_7}{g_s}\int d^8\xi\frac{\sin2\psi}{2}H\partial_{\rho}K\Phi F_{01}\  ,
\label{SPhiF}
\end{equation}
resulting in the following contribution to the complete lagrangian:\\
\begin{equation}
{\cal L}_{F\Phi}\propto \frac{\sin2\psi}{2}H\partial_{\rho}K\Phi F_{01}\  .
\label{mixing}
\end{equation}
Note that this means that only the $A_0$ and $A_1$ components of the gauge field couple to the scalar field $\Phi$. Next the contribution from the first term in (\ref{WZ}) is given by:
\begin{equation}
(2\pi\alpha')^2\frac{\mu_{7}}{g_s}\int d^8\xi\frac{(\rho^2+L_0^2)^2}{8R^4}F_{ab}F_{cd}\epsilon^{a b c d}\ ,
\end{equation}
where the indices take values along the $\rho,\psi,\alpha,\beta$ directions of the world volume. This will contribute to the equation of motion for
$A_{\rho},A_{\psi},A_{\alpha}$ and $A_{\beta}$, which do not couple to the scalar fluctuations. In this paper we will be interested in analyzing the
spectrum of the scalar modes, therefore we will not be interested in the components of the gauge field transverse to the D3-branes world volume.
However although there are no sources for these components from the scalar fluctuations, they still couple to the components along the D3-branes as
a result setting them to zero will impose constraints on the $A_{0}\dots A_3$. Indeed from the equation of motion for the gauge field along the
transverse direction one gets:
\begin{equation}
\sum\limits_{a=0}^{3}S^{aa}\partial_b\partial_a{A_a}=0,~~b=\rho,\psi,\alpha,\beta\ ,
\label{Lorenzbr}
\end{equation}
(Here, no  summation on repeated indices is intended.)
However the non-zero $B$-field explicitly breaks the Lorentz symmetry along the D3-branes' world volume. In particular we have:
\begin{eqnarray}
S^{00}=-S^{11}\ ,\quad S^{22}=S^{33}\neq S^{11}\ ,
\end{eqnarray}
which suggests that we should impose:
\begin{eqnarray}
-\partial_0{A_0}+\partial_1{A_1}=0\label{constrA}\ ,\quad\partial_2{A_2}+\partial_{3}{A_{3}}=0\ .
\end{eqnarray}
We will see that these constraints are consistent with the equations of motion for $A_{0}\dots A_3$. Indeed with this constraint the equations of
 motion for $\chi$, $\Phi$ and $A_{\mu},\mu=0\dots 3$ are,
 for $\chi$:
\begin{eqnarray}
&&\frac{1+{L'_0}^2}{g}\partial_\rho\left(\frac{g\partial_\rho\chi}{(1+{L'_0}^2)^2}\right)+\frac{\Delta_{\Omega_3}\chi}{\rho^2}
+\frac{R^4}{(\rho^2+L_0^2)^2}\widetilde{\Box}\chi+\label{EMCHI}\\
&&+\frac{1+{L'_0}^2}{g}\left(-\partial_{\rho}\left(\frac{\partial{g}}{\partial L_0}\frac{L'_0}{1+{L'_0}^2}\right)+\frac{\partial^2{g}}{\partial L_0^2}\right)\chi+\frac{1+{L'_0}^2}{g}f
F_{23}=0\ ,\nonumber
\end{eqnarray}
and for $\Phi$:
\begin{eqnarray}
\frac{1}{g}\partial_\rho\left(\frac{{g}L_0^2\partial_\rho\Phi}{1+{L'_0}^2}\right)+\frac{L_0^2\Delta_{\Omega_3}\Phi}{\rho^2}+
\frac{R^4L_0^2}{(\rho^2+L_0^2)^2}\widetilde{\Box}\Phi-\frac{H\partial_\rho
K}{g}F_{01}=0\ ,
\label{eqnPhi}
\end{eqnarray}
and  finally for $A_a$:
\begin{eqnarray}
\frac{1}{g}\partial_\rho\left(\frac{{g}\partial_\rho{A_0}}{1+{L'_0}^2}\right)+\frac{\Delta_{\Omega_3}{A_0}}{\rho^2}+
\frac{R^4}{(\rho^2+L_0^2)^2}\widetilde{\Box}{A_0}+\frac{H\partial_\rho
K}{g}\partial_1\Phi&=&0\ ,\label{EqGauge}\\
\frac{1}{g}\partial_\rho\left(\frac{{g}\partial_\rho{A_1}}{1+{L'_0}^2}\right)+\frac{\Delta_{\Omega_3}{A_1}}{\rho^2}+
\frac{R^4}{(\rho^2+L_0^2)^2}\widetilde{\Box}A_1+\frac{H\partial_\rho
K}{g}\partial_0\Phi&=&0\ ,\nonumber\\
\frac{1}{g}\partial_\rho\left(\frac{{g}\partial_\rho{A_2}}{(1+{L'_0}^2)(1+\frac{R^4H^2}{(\rho^2+L_0^2)^2})}\right)+
\frac{R^4}{(\rho^2+L_0^2)^2+R^4H^2}\widetilde{\Box}{A_2}&+&\frac{\Delta_{\Omega_3}{A_2}}{\rho^2(1+\frac{R^4H^2}{(\rho^2+L_0^2)^2})}\nonumber\\ &&\hskip2cm -\frac{f}{g}\partial_3\chi=0\ ,\nonumber\\
\frac{1}{g}\partial_\rho\left(\frac{{g}\partial_\rho{A_3}}{(1+{L'_0}^2)(1+\frac{R^4H^2}{(\rho^2+L_0^2)^2})}\right)+
\frac{R^4}{(\rho^2+L_0^2)^2+R^4H^2}\widetilde{\Box}{A_3}&+&\frac{\Delta_{\Omega_3}{A_3}}{\rho^2(1+\frac{R^4H^2}{(\rho^2+L_0^2)^2})}\nonumber\\ &&\hskip2cm+\frac{f}{g}\partial_2\chi=0\  .\nonumber
\end{eqnarray}
We have defined:
\begin{equation}
\widetilde\Box=-\partial_0^2+\partial_1^2+\frac{\partial_2^2+\partial_3^2}{1+\frac{R^4H^2}{(\rho^2+L_0^2)^2}}\ .
\end{equation}
As one can see the spectrum splits into two independent components, namely the vector modes $A_0,A_1$ couple to the scalar fluctuations along $\Phi$, while the vector modes $A_2,A_3$ couple to the scalar modes along $\chi$. However it is possible to further simplify the equations of motion for the gauge field. Focusing on the equations of motion for $A_0$ and $A_1$ in equnation~(\ref{EqGauge}), it is possible to rewrite them as:
\begin{eqnarray}
&&\frac{1}{g}\partial_\rho\left(\frac{{g}\partial_\rho{F_{01}}}{1+{L'_0}^2}\right)+\frac{\Delta_{\Omega_3}{F_{01}}}{\rho^2}+
\frac{R^4}{(\rho^2+L_0^2)^2}\widetilde{\Box}{F_{01}}-\frac{H\partial_\rho
K}{g}(-\partial_0^2+\partial_1^2)\Phi=0\label{Elec}\\
&&\frac{1}{g}\partial_\rho\left(\frac{{g}\partial_\rho{(-\partial_0{A_0}+\partial_1{A_1})}}{1+{L'_0}^2}\right)+\frac{\Delta_{\Omega_3}{(-\partial_0{A_0}+\partial_1{A_1})}}{\rho^2}+
\frac{R^4}{(\rho^2+L_0^2)^2}\widetilde{\Box}{(-\partial_0{A_0}+\partial_1{A_1})}=0\ .\nonumber
\end{eqnarray}
Note that the first constraint in (\ref{constrA}) trivially satisfies the second equation in (\ref{Elec}). In this way we are left with the first equation in (\ref{Elec}). Similarly one can show that using the second constraint in (\ref{constrA}) the equations of motion in (\ref{EqGauge}) for $A_2$ and $A_3$ boil down to a single equation for $F_{23}$:
\begin{eqnarray}
\frac{1}{g}\partial_\rho\left(\frac{{g}\partial_\rho{F_{23}}}{(1+{L'_0}^2)(1+\frac{R^4H^2}{(\rho^2+L_0^2)^2})}\right)&+&
\frac{R^4}{(\rho^2+L_0^2)^2+R^4H^2}\widetilde{\Box}{F_{23}}\nonumber\\
&+&\frac{\Delta_{\Omega_3}{F_{23}}}{\rho^2(1+\frac{R^4H^2}{(\rho^2+L_0^2)^2})}+\frac{f}{g}(\partial_2^2+\partial_3^2)\chi=0\ .
\end{eqnarray}
Now let us proceed with a study of the fluctuations along $\Phi$.
\subsection{Fluctuations Along $\Phi$}
To proceed, we have to take into account the $F_{01}$ component of the gauge field strength and solve the coupled equations of motion. Since the classical solution for the embedding of the $D$-brane is known only numerically we have to rely again on numerics to study the meson spectrum. However if we look at equation (\ref{eqnMnL}) we can see that the terms responsible for the non--trivial parts  of the equation of motion are of order $H^2$. On the other hand, the mixing of the scalar and vector modes due to the term (\ref{mixing}) appear at first order in $H$. Therefore it is possible to extract some non-trivial properties of the meson spectrum even at linear order in $H$ and as it turns out, we can observe a Zeeman--like  effect: A splitting of states that is proportional to the magnitude of the magnetic field. To describe this, let us study the approximation of weak magnetic field.
\subsubsection{Weak magnetic field}
To first order in $H$ the classical solution for the D7-brane profile is given by:
\begin{equation}
L_{0}=m+O(H^2),
\end{equation}
where $m$ is the asymptotic separation of the D3 and D7-branes and corresponds to the bare quark mass. In this approximation the expressions for $g(\rho)$ and $\partial_{\rho}K(\rho)$, become:
\begin{eqnarray}
g(\rho)=\rho^3\ ,\quad\partial_{\rho}K(\rho)=\frac{4m^2R^4\rho^3}{(\rho^2+m^2)^3}\ ,\nonumber
\end{eqnarray}
and the equations of motion for $\Phi$ and $F_{01}$, equations (\ref{eqnPhi}) and (\ref{Elec}), simplify to:
\begin{eqnarray}
&&\frac{1}{\rho^3}\left(\rho^3m^2\partial_{\rho}\Phi\right)+\frac{m^2\Delta_{\Omega_{3}}}{\rho^2}\Phi+\frac{m^2R^4}{(\rho^2+m^2)^2}\Box\Phi-4H\frac{m^2R^4}{(\rho^2+m^2)^3}F_{01}=0\ ,\\
{\rm and}\quad&&\frac{1}{\rho^3}\partial_\rho\left(\rho^3\partial_\rho F_{01}\right)+\frac{\Delta_{\Omega_3}{F_{01}}}{\rho^2}+\frac{R^4}{(\rho^2+m^2)^2}\Box{F_{01}}-4H\frac{m^2R^4}{(\rho^2+m^2)^3}{\cal P}^2\Phi=0\nonumber\ ,\\
{\rm where}\quad&&\Box=-\partial_{0}^2+\partial_{1}^2+\partial_{2}^2+\partial_{3}^2,~~~{\cal P}^2=-\partial_{0}^2+\partial_{1}^2\nonumber\ .
\label{simplified}
\end{eqnarray}
This system has become similar to the system studied in ref. \cite{Arean:2005ar} and in order to decouple it we can define the fields:
\begin{equation}
\phi_{\pm}=F_{01}\pm m{\cal P}\Phi\ ,
\end{equation}
where ${\cal P}=\sqrt{-\partial_{0}^2+\partial_{1}^2}$. The resulting equations of motion are:
 \begin{equation}
\frac{1}{\rho^3}\partial_{\rho}(\rho^3\partial_{\rho}\phi_{\pm})+\frac{\Delta_{\Omega_{3}}}{\rho^2}\phi_{\pm}+\frac{R^4}{(\rho^2+m^2)^2}\Box\phi_{\pm}\mp H\frac{4R^4m}{(\rho^2+m^2)^3}{\cal P}\phi_{\pm}=0\ .
\label{eqnMotSimpl}
\end{equation}
Note that ${\cal P}^2$ is the Casimir operator in the $(x_{0},x_{1})$ plane only, while $\Box$ is the Casimir operator along the D3-branes' world volume. If we consider a plane wave $e^{ix.k}$ then we can define:
\begin{equation}
\Box e^{ix.k}=M^2 e^{ix.k},~~~{\cal P}^2 e^{ix.k}=M_{01}^2e^{ix.k}\nonumber\ ,
\end{equation}
and we have the relation:
\begin{equation}
M^2=M_{01}^2-k_{2}^2-k_{3}^2\ .
\end{equation}
The corresponding spectrum of $M^2$ is continuous in $k_{2}, k_{3}$. However, if we restrict ourselves to motion in the $(x_{0}, x_{1})$-plane the spectrum is discrete. Indeed let us consider the ansatz:
\begin{equation}
\phi_{\pm}=\eta_{\pm}(\rho)e^{-ix_{0}k_{0}+ik_{1}x_{1}}\ .
\end{equation}
Then we can write:
 \begin{eqnarray}
&&\frac{1}{\rho^3}\partial_{\rho}(\rho^3\partial_{\rho}\eta_{\pm})+\frac{R^4}{(\rho^2+m^2)^2}M_{\pm}^2\eta_{\pm}\mp H\frac{4R^4m}{(\rho^2+m^2)^3}{M_{\pm}}\eta_{\pm}=0\ ,\label{eqn01Sm}\\
&&M_{\pm}\equiv{M_{01}}_{\pm}\nonumber\  .
\end{eqnarray}
Let us analyze equation (\ref{eqn01Sm}). It is convenient to introduce:
\begin{eqnarray}
&& y=-\frac{\rho^2}{m^2};~~~\bar M_{\pm}=\frac{R^2}{m}M_{\pm};~~~P_{\pm}(y)=(1-y)^{\alpha_{\pm}}\eta_{\pm};\label{varchan}\\
&& 2\alpha_{\pm}=1+\sqrt{1+\bar M_{\pm}^2};~~~\epsilon=H\frac{R^2}{m^2}\ .\nonumber
\end{eqnarray}
With this change of variables equation (\ref{eqn01Sm}) is equivalent to:
\begin{equation}
y(1-y)P_{\pm}''+2(1-(1-\alpha_{\pm})y)P'-\alpha_{\pm}(\alpha_{\pm-1})P_{\pm}\pm\epsilon\frac{\bar M_{\pm}}{(1-y)^2}P_{\pm}=0\ .
\label{hyper}
\end{equation}
Next we can expand:
\begin{eqnarray}
&&P_{\pm}=P_{0}\pm\epsilon P_{1}+O(\epsilon^2)\ ;~~~\alpha_{\pm}=\alpha_{0}\pm\epsilon\alpha_1+O(\epsilon^2)\ ;\label{expansions}\\
&&\bar M_{\pm}=\bar M_{0}\pm\epsilon\alpha_{1}\frac{(4\alpha_{0}+2)}{\bar M_{0}}+O(\epsilon^2)\ ;~~~\bar M_{0}=2\sqrt{\alpha_{0}(\alpha_{0}+1)} \ .\nonumber
\end{eqnarray}
leading to the following equations for $P_{0}$ and $P_{1}$:
\begin{eqnarray}
y(1-y)P_{0}''+2(1-(1-\alpha_{0})y)P_{0}'-\alpha_{0}(\alpha_{0}-1)P_{0}&=&0\ ,\label{EqPert}\\
y(1-y)P_{1}''+2(1-(1-\alpha_{0})y)P_{1}'-\alpha_{0}(\alpha_{0}-1)P_{1}&=&(\alpha_{1}(2\alpha_{0}-1)\nonumber \\ \quad &-&\frac{\bar M_{0}}{(1-y)^2})P_{0}-2\alpha_{1}y P_{0}'\ .\nonumber
\end{eqnarray}
The first equation in (\ref{EqPert}) is the hypergeometric equation and corresponds to the fluctuations in pure $AdS_{5}\times S^5$ . It has the regular solution \cite{Kruczenski:2003be}:
\begin{equation}
P_{0}(y)=F(-\alpha_{0},1-\alpha_{0},2,y)\ .
\end{equation}
Furthermore regularity of the solution for $\eta(\rho)$ at infinity requires \cite{Kruczenski:2003be} that $\alpha_{0}$  be discrete, and hence the spectrum of $\bar M_{0}$:
\begin{eqnarray}
 &&1-\alpha_{0}=-n,~~~n=0,1,\dots\label{degen}\\
 &&\bar M_{0}=2\sqrt{(n+1)(n+2)}\nonumber\ .
\end{eqnarray}
The second equation in (\ref{EqPert}) is an inhomogeneous hypergeometric equation. However for the ground state, namely $n=0$, $P_{0}=F(-1,0,2,y)=1$ and one can easily get the solution:
\begin{equation}
P_{1}(y)=\frac{\bar M_{0}}{6}\ln(1-y)+(6\alpha_{1}-\bar M_{0})(\ln(-y)+\frac{1}{y})-\frac{\bar M_{0}}{4(1-y)}\ .
\end{equation}
On the other hand, using the definition of $P_{\pm}(y)$ in (\ref{varchan})  to first order in $\epsilon$ we can write:
\begin{equation}
\eta_{\pm}=\frac{1}{(1-y)^{\alpha_{0}}}\left(1\mp\epsilon\frac{\alpha_{1}}{\alpha_0}\ln(1-y)\right)\left(1\pm\epsilon P_{1}(y)\right)\ ,
\end{equation}
for the ground state $\alpha_{0}=1$ and we end up with the following expression for $\eta_{\pm}$:
\begin{equation}
\eta_{\pm}=\frac{1}{1-y}\pm\epsilon\frac{\bar M_{0}}{4(1-y)^2}\pm\frac{\epsilon}{1-y}(6\alpha_1-\bar M_0)\left(\ln(-y)+\frac{1}{y}-\frac{\ln(1-y)}{6}\right)\ .
\label{groundstate}
\end{equation}
Now if we require that our solution is regular at $y=0$ and goes as $1/\rho^2\propto1/y $ at infinity, the last term in (\ref{groundstate}) must vanish. Therefore we have:
\begin{equation}
\alpha_{1}=\frac{\bar M_0}{6}\ .
\label{correction}
\end{equation}
\begin{figure}[h] 
   \centering
   \includegraphics[width=11cm]{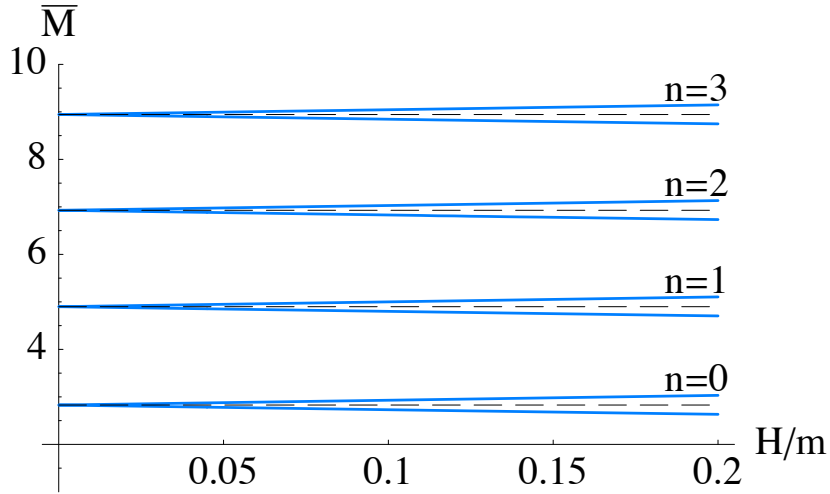}
   \caption{\small Plot of $\bar M=M{R^2}/{m}$ vs. $H/m$ for the first three states. The dashed black lines correspond to the spectrum given by equation~(\ref{degen})}
   \label{fig: Zeeman}
\end{figure}
After substituting in (\ref{expansions}) and (\ref{varchan}) we end up with the following correction to the ground sate:
\begin{equation}
M_{\pm}=M_{0}\pm\frac{H}{m}\ .
\label{Zeeman}
\end{equation}
We observe how the introduction of an external magnetic field breaks the degeneracy of the spectrum given by equation (\ref{degen}) and results in Zeeman splitting of the energy states, proportional to the magnitude of $H$. Although equation (\ref{Zeeman}) was derived using the ground state it is natural to expect that the same effect takes place for higher excited states. To demonstrate this it is more convenient to employ numerical techniques for solving equation (\ref{eqn01Sm}) and use the methods described in ref. \cite{Babington:2003vm} to extract the spectrum. The resulting plot is presented in figure \ref{fig: Zeeman}. As expected we observe Zeeman splitting of the higher excited states. It is interesting that equation (\ref{Zeeman}) describes well not only the ground state, but also the first several excited states.

It turns ou that one can easily generalize equation (\ref{Zeeman}) to the case of non-zero momentum in the $(x_{2},x_{3})$-plane. Indeed if we start from equation (\ref{eqnMotSimpl}) and proceed with the ansatz:
\begin{equation}
\phi_{\pm}=\tilde {\eta}_{\pm}(\rho)e^{-ix.k}\ ,
\end{equation}
we end up with:
\begin{eqnarray}
&&\frac{1}{\rho^3}\partial_{\rho}(\rho^3\partial_{\rho}\tilde\eta_{\pm})+\frac{R^4}{(\rho^2+m^2)^2}M_{\pm}^2\tilde\eta_{\pm}\mp H\frac{4R^4m}{(\rho^2+m^2)^3}{{M_{01}}_{\pm}}\tilde\eta_{\pm}=0\ ,\label{eqn01Sm+mom}\\
&&{M_{01}}_{\pm}=\sqrt{M_{\pm}^2+k_{23}^2};~~~k_{23}\equiv\sqrt{k_2^2+k_3^2}\ .\nonumber
\end{eqnarray}
After going through the steps described in equations (\ref{varchan})-(\ref{groundstate}), equation (\ref{correction}) gets modified to:
\begin{equation}
\alpha_{1}=\frac{\bar M_0}{6}\sqrt{1+\frac{k_{23}^2}{M_0^2}}\ .
\end{equation}
Note that validity of the perturbative analysis suggests that $\alpha_1$ is of the order of $\alpha_0$ and therefore we can trust the above expression as long as $k_{23}$ is of the order of $M_0$. Now it is straightforward to obtain the correction to the spectrum:
\begin{equation}
M_{\pm}=M_{0}\pm\frac{H}{m}\sqrt{1+\frac{k_{23}^2}{M_0^2}}\ .
\label{ZeemanGen}
\end{equation}
We see that the addition of momentum along the $(x_{2}-x_{3})$-plane enhances the splitting of the states. Furthermore the spectrum depends continuously on $k_{23}$.

\subsubsection{Strong Magnetic Field}

\begin{figure}[h] 
   \centering
  \includegraphics[width=9cm]{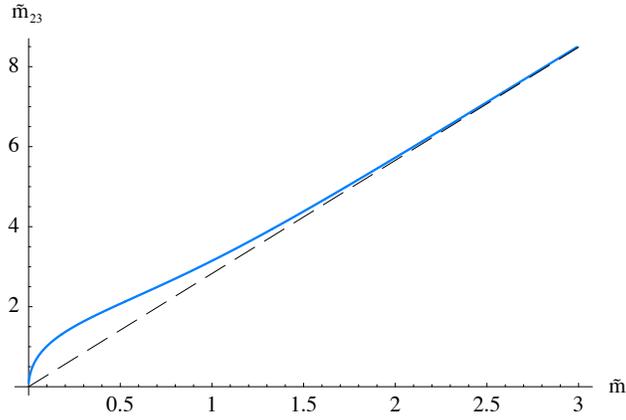}
  \caption{\small Spectrum of $\tilde m_{23}$ vs. $\tilde m$. The dashed line represents the lowest level of the meson spectrum
   for pure AdS$_{5}\times S^5$ space }
   \label{fig:fig4}
\end{figure}

For strong magnetic field we have to take into account terms of order $H^2$, which means that we no longer have an expression for $L_{0}(\rho)$ in a closed form and we have to rely on numerical calculations only. Furthermore there is no obvious way to decouple equations (\ref{eqnPhi}) and~(\ref{Elec}). However, it is still possible to extract information about the spectrum of the scalar modes if we restrict ourselves to fluctuations along the $(x^2, x^3)$ plane. In this way there is no source term in equation (\ref{EqGauge}), and we can consistently set $F_{01}$ equal to zero. We consider time independent fluctuations satisfying the ansatz $e^{-m_{23}r_{23}}$, (where $r_{23}$ is the radial coordinate in the $(x_2-x_3)$-plane). The damping factor in the exponent can be thought of as the mass of the scalar meson in 2 Euclidean dimensions. Indeed let us consider the ansatz:
\begin{equation}
\Phi=h(\rho)e^{-ik_2x^2-ik_3x^3}Y_l(S^3)\ ,
\end{equation}
where $Y_{l}(S^3)$ are the spherical harmonics on the $S^3$ sphere satisfying: $\Delta_{\Omega_3}Y_l=-l(l+2)Y_l$.  With this set-up the equation of motion for $\Phi$, equation (\ref{eqnPhi}), reduces to equation for $h(\rho)$:
\begin{eqnarray}
\frac{1}{g}\partial_\rho\left(\frac{{g}L_0^2\partial_\rho h(\rho)}{1+{L'_0}^2}\right)-\frac{L_0^2 l(l+2)}{\rho^2}h(\rho)+
\frac{R^4L_0^2m_{23}^2}{(\rho^2+L_0^2)^2+R^4H^2}h(\rho)=0\ ,
\label{eqnPhiN}
\end{eqnarray}
where we have defined:
\begin{equation}
m_{23}^2=-k_2^2-k_3^2\ .
\end{equation}
Before we proceed with the numerical analysis of equation (\ref{eqnPhiN}) let us introduce dimensionless variables by performing the transformation (\ref{cordchange}) and defining:
\begin{equation}
\tilde m_{23} =\frac{R}{\sqrt{H}}m_{23}\ .
\label{m23}
\end{equation}
The resulting equation is:
\begin{eqnarray}
&&\frac{\tilde\rho^2+\tilde L^2}{{\tilde\rho}^3\sqrt{1+\tilde{L'}^2}(1+({\tilde\rho}^2+\tilde L^2)^2)^{1/2}}\partial_{\tilde\rho}\left(\tilde\rho^3\left(1+\frac{1}{(\tilde\rho^2+\tilde L^2)^2}\right)^{1/2}\frac{\tilde L^2}{\sqrt{1+L'^2}}\partial_{\tilde\rho}h(\tilde\rho)\right)+\nonumber\\
&-&\frac{\tilde L^2}{\tilde\rho^2}l(l+2)h(\tilde\rho)+\frac{\tilde L^2\tilde m_{23}^2}{(\tilde\rho^2+\tilde L^2)^2+1}h(\tilde\rho)=0\ .
\label{eqnPhiD}
\end{eqnarray}
In order to study the spectrum we look for normalizable solutions which have asymptotic behavior $h(\tilde\rho)\propto 1/\tilde\rho^2$ for large $\tilde\rho$ and satisfy the following boundary conditions at $\tilde\rho=0$:
\begin{equation}
h'(0)=0;~~h(0)=1\ .
\label{B.C}
\end{equation}
Let us consider first the lowest level of the spectrum. The spectrum that we get as a function of the bare quark mass is plotted in figure \ref{fig:fig4}.

\begin{figure}[h] 
   \centering
  \includegraphics[width=9cm]{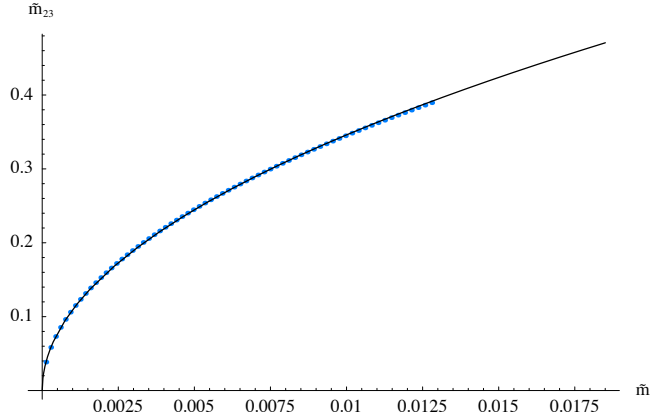}
   \caption{\small Enlargement of part of the spectrum of $\tilde m_{23}$ vs $\tilde m$ from figure~\ref{fig:fig4}. The black solid curve shows the  $\propto\sqrt{\tilde m}$ fit.}
   \label{fig:Mparal-zoomed}
\end{figure}

 \begin{figure}[h] 
   \centering
  \includegraphics[width=9cm]{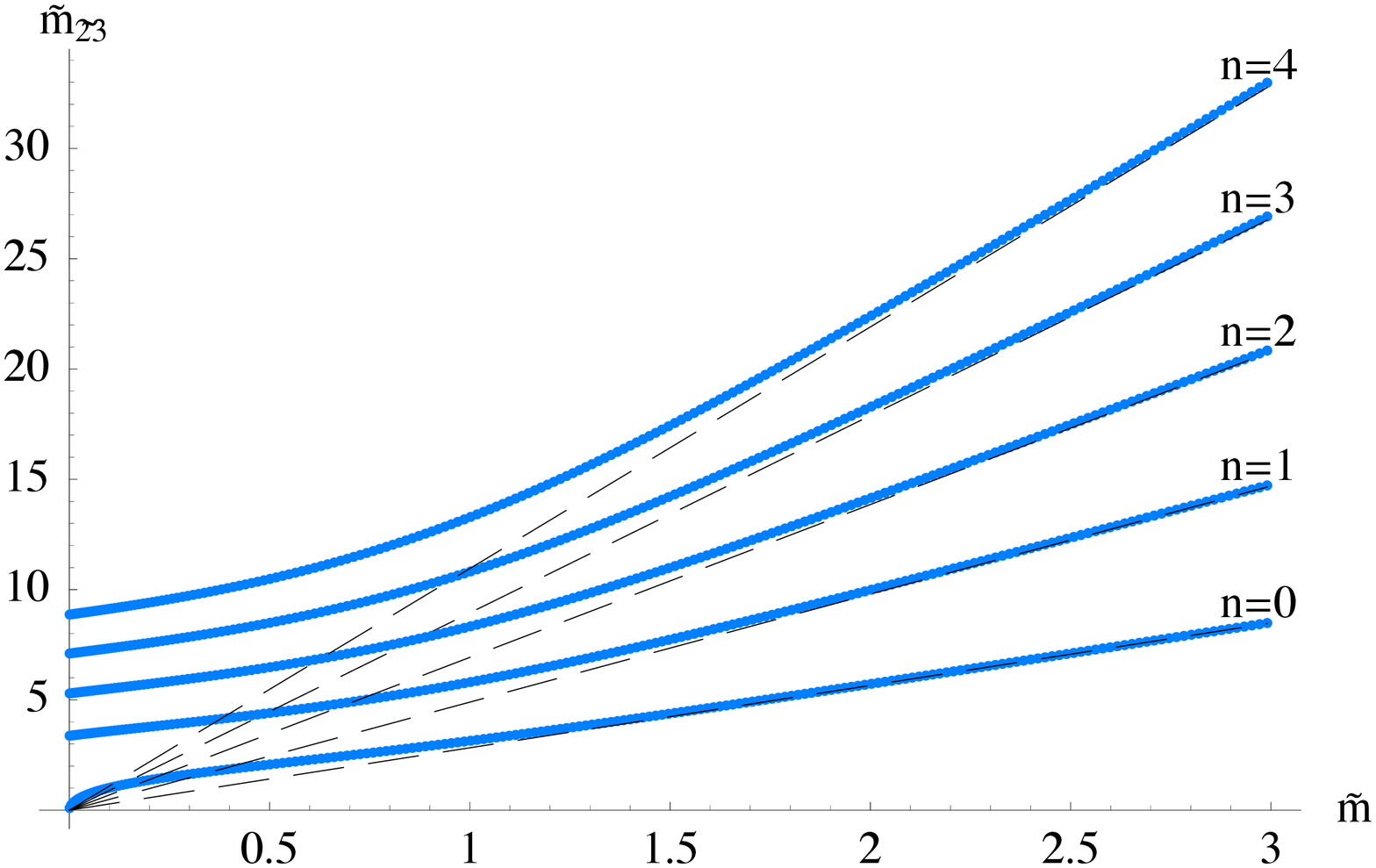}
   \caption{\small Spectrum of $\tilde m_{23}$ vs $\tilde m$ for $n=0\dots 4$. The dashed lines represent the spectrum for $AdS_5\times S^5$ space.}
   \label{fig:m23-n0-n4}
\end{figure}

For large $\tilde m$ the spectrum asymptotes (the dashed line in figure \ref{fig:fig4}) to the one  for pure $AdS_{5}\times S^5$ space obtained
by ref. \cite{Kruczenski:2003be}
\begin{equation}
M_0=\frac{2m}{R^2}\sqrt{(n+l+1)(n+l+2)}\ ,
\label{spectrAds}
\end{equation}
with the substitution $n=0, l=0$, to obtain our case. Therefore we are describing the lowest possible state of the meson spectrum. In figure {\ref{fig:Mparal-zoomed}} we have zoomed in the area near the origin of the
 $(\tilde m,\tilde m_{23})$-plane, one can see that for small values of $\tilde m=2\pi\alpha' m_q/R\sqrt{H}$ we observe $\propto\sqrt{m_{q}}$ dependence of the ground state on the bare quark mass $m_q$, which is to be expected since the chiral symmetry associated with the spinor representation of SO(2) is spontaneously broken \cite{Gell-Mann:1968rz}.

It is interesting to look for modes corresponding to higher excited states (non--zero $n$). In figure \ref{fig:m23-n0-n4} we have presented a plot of some of these. Again, the dashed line correspond to the pure $AdS_{5}\times S^5$ spectrum given by (\ref{spectrAds}) for $l=0$.  For small values of $\tilde m$ one can observe the qualitative difference of the behavior of the spectrum corresponding to the $n=0$ state from that of the higher excited states. Indeed as $\tilde m\to 0 $ the $n=0$  states follow the $\sqrt{\tilde m}$ behavior plotted in figure \ref{fig:Mparal-zoomed}, while the excited states tend to some finite values at zero bare quark mass. The $n=0$ states merge into the Goldstone boson of the spontaneously broken chiral symmetry.

\section{Acknowledgments}
V.G. Filev would like to thank T. Albash, I. Bars, A. Kundu, R. Myers and N. Warner for useful discussions and comments. The research of C. V. Johnson and V. G. Filev was supported in part by the US Department of Energy. We would specially like to thank Rene Meyer and Karl Landsteiner for commenting on the first version of the paper. The research of K.S. Viswanatan and R.C. Rashkov has been partially supported by an operating grant  from NSERC and Bulgarian NSF BUF-14/06. R.C. Rashkov and V.G. Filev thank Department of Physics and IRMACS for hospitality at the final stage of this project.

\end{document}